\documentclass[%
reprint,
superscriptaddress,
amsmath,
amssymb,
aps,
]{revtex4-2}

\usepackage{graphicx}
\usepackage{dcolumn}
\usepackage{bm}
\usepackage{braket}
\usepackage[scr=boondox]{mathalpha}

\begin{document}

\preprint{APS/123-QED}

\title{Spatiotemporal spherical Gaussian pulses}

\author{A. Konzel}
\email{akon@lle.rochester.edu}
\affiliation{Laboratory for Laser Energetics, University of Rochester, Rochester, New York 14623-1299, USA}

\author{R. Almeida}
\affiliation{GoLP/Instituto de Plasmas e Fusão Nuclear, Instituto Superior Técnico, Universidade de Lisboa, Lisbon, Portugal
}

\author{J. Vieira}
\affiliation{GoLP/Instituto de Plasmas e Fusão Nuclear, Instituto Superior Técnico, Universidade de Lisboa, Lisbon, Portugal
}

\author{J.P. Palastro}
\email{jpal@lle.rochester.edu}
\affiliation{Laboratory for Laser Energetics, University of Rochester, Rochester, New York 14623-1299, USA}

\date{\today}

\begin{abstract}
We describe a class of electromagnetic Gaussian modes with spherical spatiotemporal structure that remains invariant during propagation through anomalously dispersive media. The angular profile of the modes is given by spherical harmonics, whose quantization axis can be freely chosen. As a result, the modes support spatiotemporal orbital angular momentum (OAM) configurations with longitudinal and transverse components set by the orientation of the quantization axis. Superpositions of these modes generalize the familiar solution of a chirped pulse that simultaneously focuses and compresses to arbitrary angular structure. The modes are derived through a series of unitary transformations that map the dispersive paraxial wave equation onto a Schr\"odinger equation with a 3D isotropic harmonic potential. These transformations facilitate the identification of conserved operators for mixed longitudinal and transverse OAM.
\end{abstract}

\maketitle

\section{\label{sec:introduction} Introduction}

Engineering the structure of optical pulses has emerged as a cornerstone of modern photonics, driven by the desire to tailor the degrees of freedom of light for information processing, sensing,  and matter manipulation \cite{Forbes2021,Willner2021,Yang2021,Piccardo2025,Forbes2025,Fang2025,Cheng2025}. A seminal development came with the work of Allen \textit{et al.}, who demonstrated that paraxial laser beams with transverse Laguerre-Gaussian (LG) structure carry longitudinal orbital angular momentum (OAM), manifested by a phase vortex in the transverse plane \cite{allen_orbital_1992}. While early investigations building on this work focused primarily on spatially structured optical fields  \cite{he_direct_1995, simpson_mechanical_1997, paterson_controlled_2001, tyler_influence_2009,  wang_terabit_2012, mafu_higher-dimensional_2013,Lavery2013, nicolas_quantum_2014,Gariepy2014,Vieira2014,vieira2016,fang_orbital_2020}, recent advances have extended the paradigm into the spatiotemporal domain \cite{Sukhorukov2005,Bliokh2012,jhajj_spatiotemporal_2016,Sainte-Marie2017,Froula2018,kondakci2019optical,hancock_free-space_2019,Palastro2020,Chong2020,hancock_mode_2021,wan_toroidal_2022,zdagkas_observation_2022,Le2024,Fan2025}. This is exemplified by spatiotemporal optical vortex (STOV) pulses \cite{Bliokh2012,jhajj_spatiotemporal_2016,hancock_free-space_2019,Chong2020,  hancock_mode_2021,Fan2025}, where the spatial and temporal degrees of freedom are coupled to generate a phase vortex in a spatiotemporal plane with an associated transverse OAM.

A common formulation of STOV pulses expresses their spatiotemporal structure in terms of Laguerre-Gaussian modes in direct analogy with the transverse spatial structure introduced by Allen \textit{et al.} \cite{allen_orbital_1992}. This analogy, however, only extends so far. The transverse intensity profile of a longitudinal OAM pulse is preserved up to a rescaling during vacuum propagation. In contrast, the spatiotemporal profile of a Laguerre-Gaussian STOV pulse evolves in vacuum \cite{hancock_free-space_2019,hancock_mode_2021}.  
This evolution arises because diffraction acts on the transverse pulse structure without a corresponding effect on the temporal structure.
Preserving the intensity structure of a STOV pulse therefore requires a medium where the spatial evolution from diffraction can be balanced with the temporal evolution from dispersion. Moreover, while there is general agreement on the definition of longitudinal OAM, there has been ongoing debate over whether spatiotemporal OAM should be defined relative to the ``photon centroid" or the energy centroid of the pulse, with the two definitions yielding different values for the intrinsic and extrinsic OAM \cite{hancock_mode_2021,Bliokh2021,Hancock2024,Porras2024,Tripathi2025,Titouan2025,Vo2026}. 

Regardless of the relative effects of diffraction and dispersion and the definition of spatiotemporal OAM, the spatiotemporal phase vortex is a robust feature that persists during linear propagation. For Laguerre-Gaussian STOV pulses, the associated phase topology is restricted to two dimensions. This restriction can be overcome through more intricate spectral representations \cite{adams_spatiotemporal_2024} or by expanding the angular structure in spherical harmonics, enabling fully three-dimensional phase topologies. The latter approach was recently demonstrated experimentally by Cao \textit{et al.} \cite{cao_spatiotemporal_2025}, who described the field profile using spherical harmonics and radial Bessel functions.

\begin{figure*}
    \includegraphics[width=\textwidth]{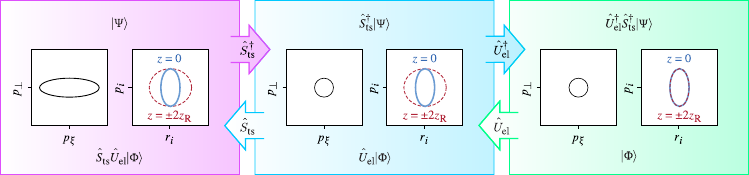}
    \caption{\label{fig:chart} Schematic of the theoretical approach, depicting the effect of the unitary transformations applied to the dispersive paraxial Hamiltonian, $\hat{H}_\mathrm{dp}$, on $1/e$ contours of the lowest order STSG pulse. The left (pink) panel illustrates the asymmetry of the pulse in momentum space and the evolution of its phase-space contour during propagation due to diffraction and dispersion. The center (blue) panel shows the result of the transformation by $\hat{S}_\mathrm{ts}^\dag$ [Eq.~\eqref{eq:S_ts}], which symmetrizes the momentum-space contour. The right (green) panel shows the subsequent transformation by $\hat{U}_\mathrm{el}^\dag$, which eliminates the evolution of the phase-space contour due to diffraction and dispersion. The unitary operators relating the states $\ket{\Psi}$ in the original system and $\ket{\Phi}$ in the fully transformed system are displayed above and below the panels; the expressions associated with each panel are equivalent representations of the same state. The subscript $i$ denotes the three coordinates $(x, y, \xi)$. }
\end{figure*}

Here, we introduce a class of Gaussian-mode pulses with three-dimensional phase topology that support OAM configurations ranging from purely longitudinal to purely spatiotemporal. These spatiotemporal spherical Gaussian (STSG) pulses preserve their structure up to a rescaling during propagation through an anomalously dispersive medium. In contrast to spherical Bessel functions, the STSG modes have finite energy individually. The lowest order mode, which has no angular structure, describes a chirped optical pulse that simultaneously focuses and compresses, reaching its minimum spot size and duration at the same location along the propagation axis. The modes are derived using a series of unitary transformations that map the dispersive paraxial wave equation onto a Schr\"odinger equation with a three-dimensional spatiotemporally symmetric harmonic potential (Fig.~\ref{fig:chart}). The unitary transformations provide a straightforward procedure for determining the OAM-like eigenoperators and eigenvalues of the STSG modes. More generally, the modes provide a convenient basis for studying the propagation and interaction of pulses with arbitrary OAM in the wide range of anomalously dispersive media.

\section{\label{sec:derivation} Derivation of STSG Pulses}

Consider a linearly polarized pulse propagating along the positive $z$-axis through a medium with a frequency-dependent refractive index $n(\omega)$. The transverse electric field of the pulse can be expressed as 
\begin{equation}
\mathbf{E}(\mathbf{r},t) = \tfrac{1}{2}\Psi(\mathbf{r},t) e^{i (k_0 z - \omega_0 t)} \, \mathbf{e}_x + \mathrm{c.c.},
\end{equation}
where $\Psi$ is the complex, slowly varying envelope, $\omega_0$ is the central frequency of the pulse, $k_0 \equiv k(\omega_0) = \omega_0n(\omega_0)/c$, and $\mathbf{e}_x$ is the unit vector in the $x$-direction. The envelope evolves according to the dispersive paraxial wave equation
\begin{equation}\label{eq:dispparax}
    i\frac{\partial \Psi}{\partial z} = -\frac{1}{2 k_0}\left(\frac{\partial^2}{\partial x^2} + \frac{\partial^2}{\partial y^2} - \beta_2 \frac{\partial^2}{\partial \xi^2}\right)\Psi,
\end{equation}
where $\beta_2 = v_g^2 k_0 k''(\omega_0)$ is the dimensionless group velocity dispersion (GVD) of the medium at $\omega_0$, $\xi \equiv v_g t - z$ is a moving-frame coordinate traveling at the group velocity $v_g = 1/k'(\omega_0)$, and primes denote frequency derivatives.

The dispersive paraxial wave equation [Eq. \eqref{eq:dispparax}] has the form of a three-dimensional Schr\"odinger equation, with $z$ playing the role of time and $\mathbf{r} \equiv (x, y, \xi)$ playing the role of the position vector. The corresponding Hamiltonian is given by
\begin{equation}\label{eq:H_dp}
    \hat{H}_\mathrm{dp} \equiv \frac{1}{2 k_0}\Bigr(\hat{p}_x^2 + \hat{p}_y^2 -\beta_2 \hat{p}_\xi^2\Bigr),
\end{equation}
where hat denotes an operator and $\hat{\mathbf{p}} \equiv (\hat{p}_x, \hat{p}_y, \hat{p}_\xi) = -i(\partial_x, \partial_y, \partial_\xi)$ is the conjugate momentum operator associated with the coordinate operator $\hat{\mathbf{r}} =(\hat{x}, \hat{y}, \hat{\xi})$. The coordinate and momentum operators obey the canonical commutation relations $[\hat{r}_i, \hat{p}_j] = i \delta_{ij}$. 

The OAM operator is defined in terms of the coordinate and momentum operators as 
\begin{equation}\label{eq:L}
    \hat{\mathbf{L}} \equiv \hat{\mathbf{r}} \times \hat{\mathbf{p}}.
\end{equation}
Note that the $\xi$ component of the OAM corresponds to the familiar longitudinal OAM: $\hat{L}_\xi = \hat{x}\hat{p}_y - \hat{y}\hat{p}_x=\hat{L}_z$.
A component of the OAM is conserved if it commutes with the Hamiltonian. To evaluate the commutation relations, it is convenient to express the Hamiltonian in the form $\hat{H}_\mathrm{dp} = 1/(2 k_0) [\hat{p}^2 - (\beta_2 + 1)\hat{p}_\xi^2]$, where $\hat{p}^2 = \hat{\mathbf{p}}\cdot\hat{\mathbf{p}}$. The first term is rotationally invariant in three dimensions, whereas the second term breaks rotational invariance about the $x$ and $y$ axes. As a result, the longitudinal OAM is conserved, $[\hat{H}_\mathrm{dp},\hat{L}_\xi] = 0$, but the total OAM is not, $[\hat{H}_\mathrm{dp}, \hat{L}^2] \propto (\beta_2 + 1)[\hat{p}_\xi^2, \hat{L}^2] \neq 0$. The exception occurs for $\beta_2 = -1$, for which the Hamiltonian becomes rotationally invariant in all three dimensions. 

The following sections derive finite-energy solutions to Eq.~\eqref{eq:dispparax} and a conserved spatiotemporal ``OAM-like'' operator for arbitrary $\beta_2 < 0$, corresponding to an anomalously dispersive medium. This is achieved through a sequence of unitary transformations that first render the original Hamiltonian (Fig. \ref{fig:chart}, left, pink) rotationally invariant in three dimensions (middle, blue) and then map it onto a Hamiltonian with a three-dimensional spatiotemporal harmonic potential (right, green). The solutions and conserved spatiotemporal OAM of the transformed Hamiltonian are then mapped back to the original system by applying the inverse unitary transformations.
 
\subsection{\label{subsec:squeeze} Temporal Squeezing Transformation}

The dispersive paraxial Hamiltonian $\hat{H}_\mathrm{dp}$ can be made rotationally invariant by applying a temporal squeezing transformation using the unitary operator $\hat{S}_\mathrm{ts}$. The transformation has the effect of rescaling or ``squeezing'' the temporal conjugate pair as $\hat{S}_\mathrm{ts}^\dag \hat{\xi} \hat{S}_\mathrm{ts} = (-\beta_2)^{1/2} \hat{\xi}$ and $\hat{S}_\mathrm{ts}^\dag \hat{p}_\xi \hat{S}_\mathrm{ts} = (-\beta_2)^{-1/2} \hat{p}_\xi$, where dagger denotes a Hermitian conjugate. The form of the operator that generates this tranformation is
\begin{equation}\label{eq:S_ts}
    \hat{S}_\mathrm{ts} \equiv \exp\left[-\frac{i}{2}\ln{[(-\beta_2)^{1/2}]}\Bigl(\hat{\xi}\hat{p}_\xi + \hat{p}_\xi \hat{\xi}\Bigr)\right].
\end{equation}
In order for $\hat{S}_\mathrm{ts}$ to be unitary, its exponent must be anti-Hermitian. This requires $\beta_2 < 0$, as is the case for anomalously dispersive media. Taking $\beta_2 = -|\beta_2|$ and applying the transformation produces the symmetrized Hamiltonian
\begin{equation}\label{eq:H_sym}
    \hat{H}_\mathrm{sym} \equiv \hat{S}_\mathrm{ts}^\dag \hat{H}_\mathrm{dp} \hat{S}_\mathrm{ts} = \frac{1}{2 k_0}\hat{p}^2.
\end{equation}
Since $\big[\hat{H}_\mathrm{sym}, \hat{\mathbf{L}}\big] = 1/(2 k_0)\big[\hat{p}^2, \hat{\mathbf{L}}\big] = 0$, all components of the OAM are conserved in this system.

\subsection{\label{subsec:EAtrans} Ermakov-Lewis Transformation}

The symmetric Hamiltonian $\hat{H}_\mathrm{sym}$ [Eq. \eqref{eq:H_sym}] depends only on the momentum magnitude $\hat{p}$. Thus, the eigenmodes of $\hat{H}_\mathrm{sym}$ are also eigenmodes of $\hat{p}$. The eigenvalues of $\hat{p}$ form a continuous spectrum, with corresponding spatial profiles described by spherical Bessel functions that extend infinitely in radius \cite{cao_spatiotemporal_2025}. Consequently, the eigenmodes are not normalizable and therefore cannot individually represent finite-energy optical pulses. Physically realizable pulses must instead be constructed from a superposition of the momentum eigenmodes. 

Discrete, finite-energy modes can be obtained by introducing a potential into $\hat{H}_\mathrm{sym}$. A harmonic potential, in particular, is known to produce solutions with a Gaussian profile, as exemplified by Hermite- and Laguerre-Gaussian  modes. The Ermakov-Lewis transformation \cite{lewis_classical_1967, lewis_class_1968} provides a systematic procedure for mapping a Schr\"odinger-like equation to one containing a harmonic potential through a unitary transformation. Here, a variation of the transformation is implemented as a sequence of two unitary transformations and a reparameterization of $z$ \cite{guerrero_quantum_2013, huerta-sandoval_unitary_2026, morales_rodriguez_spatial-light-mode_2024}. The total unitary operator is $\hat{U}_\mathrm{el} = \hat{S}_\mathrm{el}\hat{G}_\mathrm{el}$, with
\begin{subequations}
\label{eq:ELunitaries}
    \begin{eqnarray}
        \label{subeq:G_el}
        \hat{G}_\mathrm{el}(z) & \equiv & \exp\left[\frac{i}{2}k_0 \frac{\dot{w}(z)}{w(z)}\hat{r}^2\right]\\
        \label{subeq:S_el}
        \hat{S}_\mathrm{el}(z) & \equiv & \exp\left[-\frac{i}{2} \ln\left(\frac{w(z)}{w_0}\right) \left(\hat{\mathbf{r}} \cdot \hat{\mathbf{p}} + \hat{\mathbf{p}} \cdot \hat{\mathbf{r}}\right)\right],
    \end{eqnarray}
\end{subequations}
where $w(z)$ is an as-yet unspecified $z$-dependent scale factor, $w_0 \equiv w(0)$, and the overdot denotes a derivative with respect to $z$. The first operator, $\hat{G}_\mathrm{el}$, generates the desired harmonic potential but also introduces a term that couples the coordinate and momentum operators. The second operator, $\hat{S}_\mathrm{el}$, effects a $z$-dependent squeezing transformation that removes this coupling.

Under the first transformation, taking care to account for the $z$ dependence of $\hat{G}_\mathrm{el}$, the Hamiltonian becomes
\begin{equation}\label{eq_H_dp1}
\begin{split}
    \hat{G}_\mathrm{el}^\dag \hat{H}_\mathrm{sym} \hat{G}_\mathrm{el} - i \hat{G}_\mathrm{el}^\dag \dot{\hat{G}}_\mathrm{el} &= \\
    \frac{1}{2 k_0}\hat{p}^2 +  \frac{1}{2}k_0 &\frac{\ddot{w}}{w}\hat{r}^2 + \frac{1}{2}\frac{\dot{w}}{w}\left(\hat{\mathbf{r}} \cdot \hat{\mathbf{p}} + \hat{\mathbf{p}} \cdot \hat{\mathbf{r}}\right).
    \end{split}
\end{equation}
The term proportional to $\hat{r}^2$ is the desired harmonic potential. The unwanted coordinate-momentum coupling term, proportional to $\hat{\mathbf{r}} \cdot \hat{\mathbf{p}} + \hat{\mathbf{p}} \cdot \hat{\mathbf{r}}$, is removed by the isotropic squeezing operator $\hat{S}_\mathrm{el}$. The corresponding transformation rescales the conjugate coordinate and momentum pairs as $\hat{S}_\mathrm{el}^\dag \hat{\mathbf{r}} \hat{S}_\mathrm{el} = (w/w_0) \, \hat{\mathbf{r}}$ and $\hat{S}_\mathrm{el}^\dag \hat{\mathbf{p}} \hat{S}_\mathrm{el} = (w_0/w) \, \hat{\mathbf{p}}$, yielding the Hamiltonian  
\begin{equation} \label{eq:Hel}
\begin{split}    \hat{U}_\mathrm{el}^\dag \hat{H}_\mathrm{sym} \hat{U}_\mathrm{el} - i \hat{U}_\mathrm{el}^\dag \dot{\hat{U}}_\mathrm{el} &= \\ \frac{w_0^2}{w^2}\bigg[\frac{1}{2 k_0}\hat{p}^2 & + \frac{1}{2} k_0 \left(\frac{w^3 \ddot{w}}{w_0^4}\right)\hat{r}^2\bigg].
\end{split}
\end{equation}
Equation~\eqref{eq:Hel} shows that $w(z)$ determines the frequency of the harmonic potential. That is, choosing $w(z)$ to satisfy the autonomous Ermakov-Pinney equation \cite{ermakov_second-order_2008, pinney_nonlinear_1950}, $w^3\ddot{w} = w_0^4 \Omega^2$, sets the harmonic-oscillator frequency, $\Omega$. Note that, at this point, $\Omega$ is a general parameter; its relationship to physical propagation parameters will become apparent later. 

The final step to obtaining the transformed Hamiltonian is the coordinate transformation $z \rightarrow \tau$ with $d \tau /dz = w_0^2/w^2$, which eliminates the $z$-dependent coefficient of Eq.~\eqref{eq:Hel}. The resulting Hamiltonian is that of a three-dimensional isotropic harmonic oscillator, with $k_0$ playing the role of the mass:
\begin{equation}\label{eq:H_ho}
    \hat{H}_\mathrm{ho} \equiv \frac{1}{2 k_0}\hat{p}^2 + \frac{1}{2} k_0 \Omega^2 \hat{r}^2.
\end{equation}
Thus, the Ermakov-Lewis transformation defines a mapping between the symmetrized Hamiltonian $\hat{H}_\mathrm{sym}$ and the harmonic oscillator Hamiltonian $\hat{H}_\mathrm{ho}$.

\subsection{\label{subsec:oscillatorsols} Eigenmodes of the Harmonic Oscillator}

The Ermakov-Lewis transformation preserves the rotational invariance of $\hat{H}_\mathrm{sym}$. As a result, $\hat{H}_\mathrm{ho}$ admits simultaneous eigenmodes of itself, $\hat{L}^2$, and a component of $\hat{\mathbf{L}}$ along a freely chosen axis $q$ with unit vector $\mathbf{e}_q$: $\hat{L}_q \equiv \mathbf{e}_q \cdot \hat{\mathbf{L}}$. The joint eigenmodes of these operators are the familiar harmonic-oscillator modes $\ket{k \ell m}$, labeled by the radial index $k$, angular degree $\ell$, and azimuthal order $m$, where $k$ and $\ell$ are nonnegative integers and $m = -\ell, ..., \ell$. The position-space eigenfunctions associated with these modes, $\phi_{k \ell m}(\mathbf{r}) \equiv \braket{\mathbf{r}|k\ell m}$, are given in Appendix A. The $\ket{k \ell m}$ modes satisfy the eigenvalue relations
\begin{subequations}
\label{eq:qho_eigenstuff}
    \begin{eqnarray}
        \label{subeq:H_ho_eigen}
        \hat{H}_\mathrm{ho}\ket{k \ell m} & = & \Omega \big(2k + \ell + \tfrac{3}{2}\big)\ket{k \ell m}\\
        \hat{L}^2 \ket{k \ell m} & = & \ell(\ell + 1)\ket{k \ell m}\\
        \label{subeq:Lq_eigen}
        \hat{L}_q\ket{k \ell m} & = & m\ket{k \ell m}.
    \end{eqnarray}
\end{subequations}
These relations show that the $\ket{k\ell m}$ modes have an OAM magnitude of $\sqrt{\ell(\ell+1)}$ and an OAM  component $m$ along the quantization axis $q$. For completeness, the corresponding OAM ladder operators are $\hat{L}_\pm = \hat{L}_x \pm i(\mathbf{e}_q\times\mathbf{e}_x)\cdot\hat{\mathbf{L}}$, which raise and lower the azimuthal order according to $\hat{L}_\pm\ket{k \ell m} = \sqrt{\ell(\ell+1)-m(m\pm1)}\ket{k, \ell, m \pm 1}$.

\begin{figure*}
    \includegraphics[width=\textwidth]{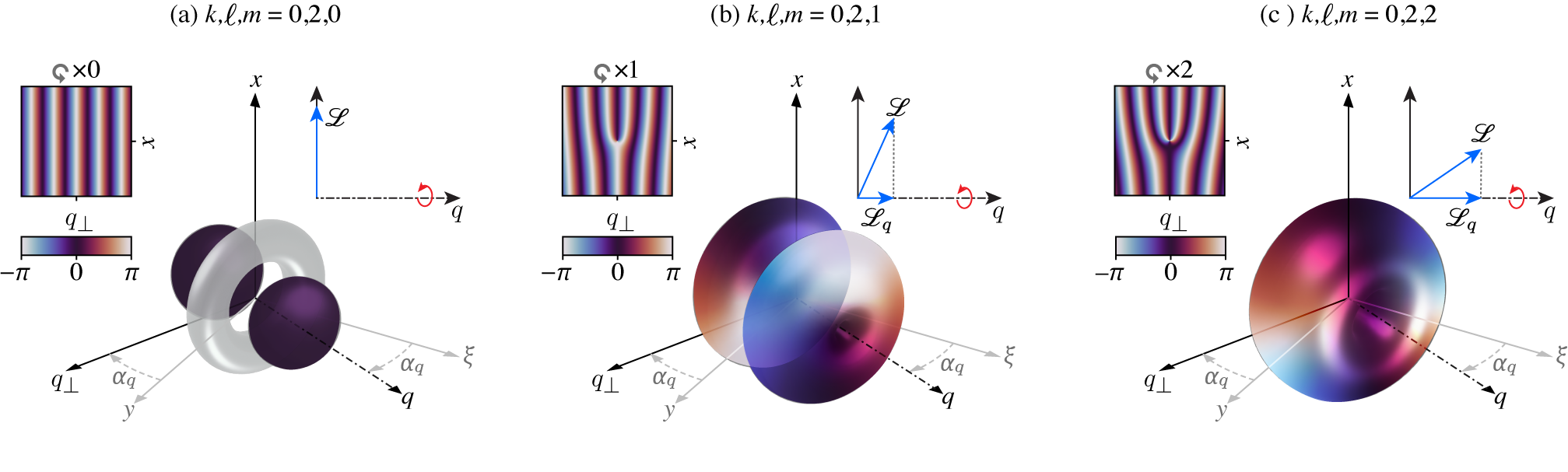}
    \caption{\label{fig:3Dplots} Phase of an STSG envelope, $\Psi_{k \ell m}(\mathbf{r})$, on its $1/e^2$ intensity isocontour for (a) $(k,\ell,m) = (0,2,0)$, (b) $(k,\ell,m)=(0,2,1)$, and (c) $(k,\ell,m)=(0,2,2)$. Here, the dimensionless GVD $\beta_2 = -1$, such that $\mathscr{q} = q$. The upper-left insets show the total phase of the transverse electric field $\mathbf{e}_\mathscr{x}\cdot\mathbf{E}$, including the carrier, in the plane spanned by  $\mathbf{e}_{q_\perp} = \mathbf{e}_q \times\mathbf{e}_x$ and $\mathbf{e}_x$. The upper-right insets show the $q$ component, $\mathscr{L}_q$, and magnitude, $\mathscr{L}$, of the OAM-like operator $\bm{\mathscr{L}}$. The circular red arrow indicates that the OAM-like contribution perpendicular to the $q$ axis has no preferred azimuthal orientation. The index $m$ determines $\mathscr{L}_q$ and the counterclockwise phase circulation in the $q_\perp-x$ plane. The three examples were chosen to illustrate  STSG pulses with (a) the minimum $\mathscr{L}_q$, (b) an intermediate $\mathscr{L}_q$, and (c) the maximum $\mathscr{L}_q$ for the given $\ell$ value. }
\end{figure*}

\subsection{\label{subsec:STSGsols} Spatiotemporal Spherical Gaussian Modes}

Consider a single harmonic-oscillator state at $\tau = 0$ given by $\ket{\Phi(0)} \equiv \ket{k \ell m}$. From Eq.~\eqref{subeq:H_ho_eigen}, the state at any other $\tau$ is given by 
\begin{equation}\label{eq:phitau}
    \ket{\Phi(\tau)} =  \exp\left[-i \Omega (2k + \ell + \tfrac{3}{2}) \tau\right] \ket{k \ell m},
\end{equation}
with the corresponding wavefunction $\Phi(\mathbf{r},\tau) = \braket{\mathbf{r}|\Phi(\tau)}$.
Mapping this state back to the original system yields the evolution of the mode $\ket{\Psi(z)}$ under the dispersive paraxial Hamiltonian $\hat{H}_\mathrm{dp}$: 
\begin{equation}\label{eq:psiz}
    \ket{\Psi(z)} = \exp[-i \Omega (2 k + \ell + \tfrac{3}{2})\tau(z)] \hat{S}_\mathrm{ts} \hat{U}_\mathrm{el}(z) \ket{k \ell m}.
\end{equation}
The position-space profiles associated with the dispersive paraxial modes, $\psi_{k\ell m}(\mathbf{r},z) \equiv\braket{\mathbf{r}|\hat{S}_\mathrm{ts} \hat{U}_\mathrm{el}(z) |k \ell m}$, are provided in Appendix A. 

Recall that the form of $\hat{H}_\mathrm{ho}$ requires $w(z)$ to satisfy $w^3 \ddot{w} = w_0^4 \Omega^2$, where the constant $\Omega$ can be freely chosen. Setting $\Omega = 1/z_\mathrm{R}$ with the initial condition $\dot{w}(0) = 0$, where $z_\mathrm{R} \equiv k_0 w_0^2/2$ is the Rayleigh range, provides a direct correspondence with the familiar evolution of a Gaussian beam. These choices and integration of $d\tau = (w_0/w)^2 dz$ yield the position-space wavefunctions of the spatiotemporal spherical Gaussian (STSG) modes, $\Psi_{ k \ell m}(\mathbf{r},z) = \braket{\mathbf{r}|\Psi(z)}$:
\begin{equation}
\begin{split}\label{eq:envelope}
    \Psi_{k \ell m}(\mathbf{r}, z) = &\frac{N_{k \ell}}{|\beta_2|^{1/4}}\frac{1}{\sqrt{w^3}}\bigg(\frac{\sqrt{2}\varrho}{w}\bigg)^\ell 
    L_k^{\ell + 1/2}\left(\frac{2 \varrho^2}{w^2}\right) \\ &\times \exp\left[-\frac{\varrho^2}{w^2} + \frac{ik_0\varrho^2 }{2R}-i \Theta_{k\ell}\right] Y_\ell^m(\vartheta, \varphi),
\end{split}
\end{equation}
where $\bm{\varrho} \equiv \hat{S}_\mathrm{ts}\mathbf{r}\hat{S}_\mathrm{ts}^\dagger = (x, y, |\beta_2|^{-1/2} \xi)$, $L_k^{\ell+1/2}$ is the generalized Laguerre polynomial, $Y_\ell^m$ is the spherical harmonic, and $(\vartheta,\varphi)$ are polar and azimuthal angles with respect to a chosen polar axis in the squeezed coordinate $\bm{\varrho}$. The propagation-dependent quantities
\begin{subequations}
\label{eq:zfuncs}
    \begin{eqnarray}
        \label{subeq:spotsize}
        w(z) &=& w_0 \sqrt{1 + (z/z_\mathrm{R})^2},\\
        \label{subeq:gouy}
        \Theta_{k \ell}(z) &=& (2 k + \ell + \tfrac{3}{2})\arctan(z/z_\mathrm{R}),\\
        \label{subeq:curvature}
        R(z) &=& z[1 + (z_\mathrm{R}/z)^2],
    \end{eqnarray}
\end{subequations}
describe the spherical width, Gouy phase, and radius of curvature of the mode, respectively. The normalization factor $N_{k \ell} \equiv 2^{3/4} \sqrt{2 k !/\Gamma(k + \ell + 3/2)}$ is chosen so that $\int |\Psi_{k \ell m}|^2 \, d^3\mathbf{r} = 1$. 

At this stage, the polar and azimuthal angles in the squeezed coordinates, $\vartheta$ and $\varphi$, can be defined relative to an arbitrary orientation of the quantization axis. To obtain explicit expressions for these angles, the quantization axis $q$ of the harmonic-oscillator system can be placed in the $y$-$\xi$ plane, which allows for OAM that can be oriented anywhere between transverse and longitudinal. Let $\alpha_q$ denote the angle between the $q$ and $\xi$ axes, such that
\begin{equation}
\mathbf{e}_q = \sin{\alpha_q}\,\mathbf{e}_y + \cos{\alpha_q}\,\mathbf{e}_\xi.
\end{equation}
The spherical coordinates of the harmonic-oscillator system, $(r,\theta,\phi)$, can then be oriented so that $\mathbf{e}_q$ defines the polar axis (see Appendix B). The corresponding temporally squeezed spherical coordinates are 
\begin{align}
        \label{subeq:varrho}
        &\varrho = \sqrt{x^2 + y^2 + |\beta_2|^{-1} \xi^2} \tag{17a} \\
        \label{subeq:vartheta}
        &\vartheta = \arccos\left(\frac{ y \sin{\alpha_q} + |\beta_2|^{-1/2}\xi\cos{\alpha_q}}{\varrho}\right) \tag{17b}\\
        \label{subeq:varphi}
        &\varphi = \arctan\left(\frac{y \cos{\alpha_q} - |\beta_2|^{-1/2}\xi \sin{\alpha_q}}{x}\right). \tag{17c}
\end{align}
Thus, the specification of $\alpha_q$ allows the quantization axis of the STSG modes to be varied continuously between purely longitudinal ($\alpha_q=0$) and purely transverse ($\alpha_q=\pi/2$) configurations. Figure \ref{fig:3Dplots} shows the phase structure of three different STSG modes, with the harmonic-oscillator quantization axis rotated in the $y-\xi$ plane by an angle $\alpha_q$ with respect to the $\xi$-axis.
\setcounter{equation}{17}

\begin{figure}
    \centering
    \includegraphics[width=0.45\textwidth]{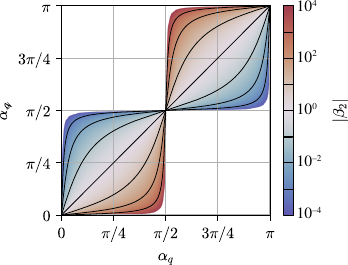}
    \caption{\label{fig:alphacomp} Relationship between the angles of the harmonic-oscillator quantization axis, $\alpha_q$, and the vortex axis, $\alpha_\mathscr{q}$, with respect to the $\xi$-axis in the $y-\xi$ plane. The two angles coincide at multiples of $\pi/2$ regardless of $\beta_2$.
    Away from multiples of $\pi/2$, the  seperation between $\alpha_q$ and $\alpha_\mathscr{q}$ increases with $|\ln|\beta_2||$. In the special case $\beta_2 = -1$, $\alpha_q = \alpha_\mathscr{q}$ for all $\alpha_q$.}
\end{figure}

When $|m|>0$, the azimuthal factor $e^{i m \varphi}$ of $Y_\ell^m(\vartheta, \varphi)$ produces a phase vortex whose core is the line along which $\varphi$ [Eq.~\eqref{subeq:varphi}] is singular. The unit vector along the vortex core is given by
\begin{equation}
\mathbf{e}_\mathscr{q} \equiv \frac{\sin{\alpha_q}\,\mathbf{e}_y + |\beta_2|^{1/2}\cos{\alpha_q}\,\mathbf{e}_\xi}{\sqrt{\sin^2{\alpha_q}+|\beta_2|\cos^2{\alpha_q}}}.
\end{equation}
Rewriting the unit vector in the form $\mathbf{e}_\mathscr{q} = \sin{\alpha_\mathscr{q}}\,\mathbf{e}_y +
\cos{\alpha_\mathscr{q}}\,\mathbf{e}_\xi$,
where
\begin{equation}\label{eq:alfq}
\alpha_\mathscr{q} = \arctan\big(|\beta_2|^{-1/2}\tan{\alpha_q}\big),
\end{equation}
reveals that the vortex axis is generally distinct from the harmonic-oscillator quantization axis $\mathbf{e}_q$ and is oriented in the $y$-$\xi$ plane at an angle $\alpha_\mathscr{q}$ with respect to the $\xi$ axis. The quantization axis and vortex axis are aligned for arbitrary $\alpha_q$ when $\beta_2 = -1$, as well as for purely longitudinal ($\alpha_q=0$) and purely transverse ($\alpha_q=\pi/2$) configurations for any $\beta_2 < 0$. The relationship between $\alpha_q$ and $\alpha_\mathscr{q}$ for varying values of $\beta_2$ is shown in Fig.~\ref{fig:alphacomp}. 

\subsection{\label{subsec:temporalwaist} STSG Pulses}

An optical pulse initialized at any $z$ with the spatiotemporal structure described by Eq.~\eqref{eq:envelope} preserves this structure, up to the rescaling described by $w(z)$, during propagation through an anomalously dispersive medium. Because the STSG modes are normalized according to $\int |\Psi_{k \ell m}|^2 \, d^3\mathbf{r} = 1$, the amplitude of the pulse is determined by its energy $U$, such that its envelope is given by
\begin{equation}
\Psi(\mathbf{r},z) = \left( \frac{2v_gU}{\epsilon_0cn(\omega_0)} \right)^{1/2} \Psi_{k \ell m}(\mathbf{r},z).
\end{equation}
The $|\beta_2|^{-1/2}$ scale factor accompanying $\xi$ in $\varrho$ [Eq.~\eqref{subeq:varrho}] implies a pulse length $w_\xi = |\beta_2|^{1/2}w$ with a corresponding duration  $v_g^{-1}|\beta_2|^{1/2}w$. As a consequence, while the STSG modes are expressed in terms of temporally squeezed spherical coordinates [Eqs.~(17)], they are only truly spherical when $\beta_2 = -1$. In addition, the quadratic phase in Eq.~\eqref{eq:envelope} reveals that an STSG pulse has a wavefront curvature and temporal chirp that are matched such that the pulse first undergoes simultaneous focusing and compression, becomes transversely and temporally transform limited at $z=0$, and then undergoes simultaneous defocusing and stretching. In this sense, $w_\xi/w = |\beta_2|^{1/2}$ can be interpreted as the condition for an STSG pulse to maintain its spatiotemporal structure. A pulse prepared with $w_\xi/w \neq |\beta_2|^{1/2}$ will not maintain this structure throughout propagation. 

\section{\label{sec:oam} Eigenoperators}

Section \ref{subsec:oscillatorsols}  established the harmonic-oscillator $\ket{k \ell m}$ modes as simultaneous eigenmodes of the total OAM operator $\hat{L}^2$ and the OAM component $\hat{L}_q = \mathbf{e}_q \cdot \hat{\mathbf{L}}$. Consequently, the STSG modes $\hat{S}_\mathrm{ts} \hat{U}_\mathrm{el} \ket{k \ell m}$ are eigenmodes of the transformed OAM operators 
\begin{equation*}
\begin{split}
\hat{\mathscr{L}}^2 \equiv \hat{S}_\mathrm{ts} \hat{U}_\mathrm{el} \hat{L}^2 \hat{U}_\mathrm{el}^\dag \hat{S}_\mathrm{ts}^\dag 
\quad \mathrm{and} \quad \hat{\mathscr{L}}_q \equiv  \hat{S}_\mathrm{ts} \hat{U}_\mathrm{el} \hat{L}_q \hat{U}_\mathrm{el}^\dag \hat{S}_\mathrm{ts}^\dag.
\end{split}
\end{equation*}
The form of the transformed operators simplifies because $\hat{U}_\mathrm{el}$ is rotationally invariant and therefore commutes with $\hat{\mathbf L}$, such that $\hat{U}_\mathrm{el} \hat{\mathbf{L}} \hat{U}_\mathrm{el}^\dag = \hat{\mathbf{L}}$. The temporally squeezed OAM operator can then be written as $\hat{\pmb{\mathscr{L}}} \equiv \hat{S}_\mathrm{ts} \hat{\mathbf{L}} \hat{S}_\mathrm{ts}^\dag$, with the scalar operators
\begin{equation*}
\begin{split}
\hat{\mathscr{L}}^2 = \hat{S}_\mathrm{ts} \hat{L}^2\hat{S}_\mathrm{ts}^\dag \quad \mathrm{and} \quad  \hat{\mathscr{L}}_q =  \hat{S}_\mathrm{ts} \hat{L}_q  \hat{S}_\mathrm{ts}^\dag. 
\end{split}
\end{equation*}
These ``OAM-like" operators have the same eigenvalues when acting on the STSG modes as the original OAM operators acting on the $\ket{k \ell m}$ modes:
\begin{subequations}
    \label{eq:Lts_eigenstuff}
    \begin{eqnarray}
        \label{subeq:L2ts_eigenstuff}
        \hat{\mathscr{L}}^2 \big(\hat{S}_\mathrm{ts} \hat{U}_\mathrm{el} \ket{k \ell m}\big) & = & \ell(\ell + 1) \big(\hat{S}_\mathrm{ts} \hat{U}_\mathrm{el} \ket{k \ell m}\big) \\
        \label{subeq:Lqts_eigenstuff}
        \hat{\mathscr{L}}_q \big(\hat{S}_\mathrm{ts} \hat{U}_\mathrm{el} \ket{k \ell m}\big) & = & m \big(\hat{S}_\mathrm{ts} \hat{U}_\mathrm{el} \ket{k \ell m}\big).
    \end{eqnarray}
\end{subequations}
Moreover, the operators and $\hat{\pmb{\mathscr{L}}}$ commute with the dispersive paraxial Hamiltonian [Eq.~\eqref{eq:H_dp}] and are conserved during propagation. The corresponding OAM-like ladder operators are $\hat{\mathscr{L}}_\pm = \hat{S}_\mathrm{ts} \hat{L}_\pm  \hat{S}_\mathrm{ts}^\dag$.

\subsection{\label{subsec:Representations} Coordinate-Space Representations}

Consistent with the spherical-harmonic dependence $Y_\ell^m(\vartheta, \varphi)$ in Eq.~\eqref{eq:envelope}, the position-space representations of the OAM-like operators take the standard spherical forms when expressed in terms of $\vartheta$ and $\varphi$:
\begin{subequations}
    \label{eq:OAMloangles}
    \begin{eqnarray}
        \label{subeq:L2ts_spherical}
        \hat{\mathscr{L}}^2 &=& -\frac{1}{\sin{\vartheta}}\frac{\partial}{\partial \vartheta}\left(\sin{\vartheta}\frac{\partial}{\partial \vartheta}\right) - \frac{1}{\sin^2{\vartheta}}\frac{\partial^2}{\partial \varphi^2} \\
        \label{subeq:Lqts_spherical}
        \hat{\mathscr{L}}_q &=& -i\frac{\partial}{\partial \varphi}.
    \end{eqnarray}
\end{subequations}
In the dispersive paraxial system, $\hat{\mathscr{L}}_q$ cannot generally be expressed as a projection of the ordinary OAM operator $\hat{\mathbf{L}}$ [Eq.~\eqref{eq:L}] along a fixed axis. Instead, $\hat{\mathscr{L}}_q$ generates rotations about the $q$ axis within the temporally squeezed system. 

The effect of the temporal squeezing can be made explicit by expressing the OAM-like operators in the original Cartesian coordinates $(x, y, \xi)$ of the dispersive paraxial system. The total OAM-like operator becomes
\begin{eqnarray}
    \label{eq:L2ts_Cartesian}
    \hat{\mathscr{L}}^2 = \hat{L}^2 + \left(1 - |\beta_2|^{-1}\right)\xi^2 \nabla_\perp^2 + (1 - |\beta_2|)r_\perp^2\frac{\partial^2}{\partial \xi^2},
\end{eqnarray}
where $r_\perp^2=x^2+y^2$ and $\nabla_\perp^2=\partial_x^2+\partial_y^2$. For $\hat{\mathscr{L}}_q$, it is instructive to first write
\begin{equation}\label{eq:Lqcombo}
\hat{\mathscr{L}}_q =
\hat{\mathscr{L}}_y\sin\alpha_q +
\hat{\mathscr{L}}_\xi\cos\alpha_q.
\end{equation}
Since $\hat{L}_\xi=\hat{x}\hat{p}_y-\hat{y}\hat{p}_x$ is independent of $\hat{\xi}$ and $\hat{p}_\xi$, it is unaffected by the temporal squeezing, so that $\hat{\mathscr{L}}_\xi = \hat{L}_\xi$:
\begin{equation} \label{eq:Lxi}
    \hat{\mathscr{L}}_\xi  =  -i\left(x\frac{\partial}{\partial y} - y\frac{\partial}{\partial x}\right).
\end{equation}
In contrast, $\hat{L}_y=\hat{\xi}\hat{p}_x-\hat{x}\hat{p}_\xi$ depends on both $\hat{\xi}$ and $\hat{p}_\xi$ and is therefore modified by the temporal squeezing:
\begin{equation}
    \label{eq:Lyts}
        \hat{\mathscr{L}}_y  =  -i\left(|\beta_2|^{-1/2} \xi \frac{\partial}{\partial x} - |\beta_2|^{1/2} x \frac{\partial}{\partial \xi}\right).
\end{equation}
When $\beta_2=-1$, the effects of the temporal squeezing vanish and the OAM-like operators reduce to their ordinary OAM counterparts, $\hat{\mathscr{L}}^2 =\hat{L}^2$, $\hat{\mathscr{L}}_y = \hat{L}_y$, and $\hat{\mathscr{L}}_q = \hat{L}_q$, consistent with the rotational symmetry of the dispersive paraxial Hamiltonian [Eq.~\eqref{eq:H_dp}].

\subsection{\label{subsec:Interpretation} Interpretation}

Several properties of the OAM-like operators can be elucidated by examining their behavior in specific limits. First, note that $\hat{\mathscr{L}}_q$ takes particularly simple forms in two configurations. In the purely longitudinal limit, $\alpha_q = \alpha_\mathscr{q} = 0$ and $\hat{\mathscr{L}}_q = \hat{L}_\xi$. Under these conditions, $\hat{\mathscr{L}}_q$ is no longer ``OAM-like''; it becomes the genuine longitudinal OAM operator. In the purely transverse limit, $\alpha_q = \alpha_\mathscr{q} = \pi/2$. Here, $\hat{\mathscr{L}}_q = \hat{\mathscr{L}}_y$, which generally cannot be expressed as a projection of $\hat{\mathbf{L}}$ along a fixed axis. This is because the temporal squeezing weights its two terms differently [Eq.~\eqref{eq:Lyts}]. For intermediate quantization angles with $\beta_2 \neq -1$, $\alpha_q \neq \alpha_\mathscr{q}$. In this case, $\hat{\mathscr{L}}_q$ is a linear combination of $\hat{L}_\xi$ and $\hat{\mathscr{L}}_y$ [Eq.~\eqref{eq:Lqcombo}] and likewise cannot be expressed as a projection of $\hat{\mathbf{L}}$ along a fixed axis.

Of particular note are STSG modes with $\ell>0$ and $m=0$. These modes have a nonzero OAM-like magnitude, $\mathscr{L}=\sqrt{\ell(\ell+1)}$, despite having no phase vortex and a vanishing OAM-like component, $\mathscr{L}_q=0$. The vanishing component $\mathscr{L}_q=0$ confines the OAM-like magnitude to the plane perpendicular to the $q$ axis, without selecting a particular orientation within that plane. The leftmost panel of Fig.~\ref{fig:3Dplots} shows an example for the mode $(k,\ell,m)=(0,2,0)$, with the inset illustrating the indeterminate orientation. For longitudinal quantization, $\alpha_q=\alpha_\mathscr q=0$, the OAM-like magnitude is therefore confined to the transverse $x-y$ plane—a configuration distinct from yet nonetheless connected to that of a conventional STOV pulse.

The $m = 0$ modes contrast with those having the maximum allowed value $|m| = \ell$. The rightmost panel of Fig.~\ref{fig:3Dplots} shows an example for the mode $(k,\ell,m) = (0,2,2)$. For modes of this type, the OAM-like component along $\mathbf{e}_q$ is maximized, while the OAM perpendicular to $\mathbf{e}_q$ is minimized. Nonetheless, $\pmb{\mathscr{L}}$ cannot be confined entirely to the $q$ direction, as an OAM normal component of magnitude $\sqrt{\ell(\ell+1)-\ell^2} =\sqrt{\ell}$ always remains.

\begin{figure}
    \centering
    \includegraphics[width=0.35\textwidth]{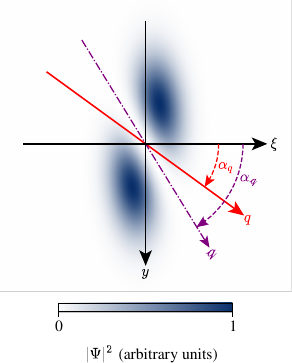}
    \caption{\label{fig:offset} The squared magnitude of an STSG pulse with $(k,\ell,m) = (0,1,1)$ in the $y-\xi$ plane at $x=0$ for  $\beta_2 = -0.2$ and $\alpha_q = \pi/5$. The phase vortex is oriented along the $\mathscr{q}$ axis at an angle $\alpha_\mathscr{q}$ with respect to the $\xi$ axis, while the $q$ axis defining the OAM-like component $\hat{\mathscr{L}}_q$ is oriented at the angle $\alpha_q$. }
\end{figure}

A final consequence concerns the relationship between the two axes associated with STSG modes: the $q$ axis defining the component $\mathscr{L}_q$ of the OAM-like operator $\hat{\pmb{\mathscr{L}}}$, and the $\mathscr{q}$ axis of the phase vortex embedded in the STSG envelope. While both axes lie in the $y-\xi$ plane, the temporal squeezing generally causes them to be oriented at different angles with respect to the $\xi$ axis. The relationship between these angles is given by Eq.~\eqref{eq:alfq}. Figure \ref{fig:offset} illustrates their relative orientation for an STSG pulse with $(k,\ell,m) = (0,1,1)$. As discussed earlier, the axes coincide only in the special cases of purely longitudinal ($\alpha_q = 0$) or purely transverse ($\alpha_q = \pi/2$) quantization, or for all $\alpha_q$ when $\beta_2 = -1$.

\section{\label{sec:conclusion} Conclusion and Discussion}

Spatiotemporal spherical Gaussian (STSG) pulses support spatiotemporal OAM configurations ranging from longitudinal to transverse and maintain their structure during propagation through an anomalously dispersive medium. In doing so, STSGs unify these forms of OAM within a single optical structure and extend the associated two-dimensional phase topologies to three dimensions. To maintain their structure, STSG pulses simultaneously focus and compress, and thus represent a generalization of the well-known spherically symmetric three-dimensional focusing pulses to arbitrary values of OAM. As a direct consequence of their three-dimensional structure, STSG pulses can carry OAM without exhibiting a phase vortex (i.e., when $\ell>0$ and $m=0$).

The STSG modes were derived by mapping the dispersive paraxial wave equation to a Schr\"odinger equation with a spatiotemporally isotropic harmonic potential. The mapping involves a sequence of unitary transformations, beginning with temporal squeezing and followed by the Ermakov-Lewis transformation. The quantization axis of the harmonic oscillator system ($q$) can be chosen arbitrarily, yielding spherical eigenmodes that map back to the STSG modes of the original dispersive paraxial system. The resulting solutions reveal that the pulse length and spot size of the STSG pulses must satisfy the matching condition, $w_\xi/w = |\beta_2|^{1/2}$, which also determines their eccentricity. 

The correspondence between the STSG modes and the harmonic oscillator OAM eigenstates allows for the identification of ``OAM-like'' operators $\hat{\mathscr{L}}^2$ and $\hat{\mathscr{L}}_q$ that are conserved during propagation. As a result of the temporal squeezing, the $q$ axis defining the conserved component $\hat{\mathscr{L}}_q$ is generally misaligned with the vortex axis. The two axes coincide only in special cases, such as the spherically symmetric case where $\beta_2 = -1$, or for purely longitudinal or transverse OAM.

While this work has focused on STSG modes, the operator $\hat{\mathscr{L}}_q$ [Eq.~\eqref{eq:Lqcombo}] also applies to Laguerre-Gaussian modes. Such modes may be obtained by a procedure analogous to that presented here, except that cylindrical rather than spherical eigenmodes of the harmonic oscillator Hamiltonian are chosen. As done here, the quantization axis $q$ can be restricted to the $y-\xi$ plane and oriented at an angle $\alpha_q$ with respect to the $\xi$ axis. When $\alpha_q = \pi/2$, the resulting solutions are Laguerre-Gaussian STOVs with OAM-like operator $\hat{\mathscr{L}}_y$ [Eq.~\eqref{eq:Lyts}], whereas when $\alpha_q = 0$, they are standard longitudinal-OAM Laguerre-Gaussian modes with OAM operator $\hat{L}_\xi$ [Eq.~\eqref{eq:Lxi}].

An optical pulse constructed without the matched eccentricity $w_\xi/w = |\beta_2|^{1/2}$, or one propagating in a medium with $\beta_2 \geq 0$, will not preserve its structure during propagation. Nevertheless, its evolution can still be described analytically. For instance, the Ermakov-Lewis transformation can be applied independently to each direction, with distinct spot sizes $w_x$, $w_y$ and duration $w_\xi$. The resulting Hamiltonian would have an anisotropic harmonic potential, but would remain separable in the Cartesian coordinates $(x, y, \xi)$, allowing a general state to be constructed from products of independent oscillator modes. Upon transforming back to the original system, these modes correspond to pulses with Hermite-Gaussian structure. Following an approach similar to that of Ref.~\cite{hancock_mode_2021}, suitable superpositions of these modes could recover STSG or Laguerre-Gaussian pulses in the matched case, or describe the evolution of pulses initialized in these forms when the matching condition is not satisfied. This construction is generally more involved than that for the matched case considered here and is left for future work.

\appendix

\section{Position-Space Profile of  the Eigenmodes}

The $\ket{k \ell m}$ eigenmodes of the isotropic harmonic oscillator, with the eigenvalues in Eq.~\eqref{eq:qho_eigenstuff}, have position-space profiles $\phi_{k\ell m}(\mathbf{r}) \equiv \left<\mathbf{r}|k \ell m\right>$ given by
\begin{equation}
    \label{eq:phiklm_wavefunc}
    \begin{split}
        \phi_{k\ell m}(\mathbf{r}) = N_{k \ell} & \left(\frac{k_0 \Omega}{2}\right)^{3/4} \big(\sqrt{k_0 \Omega}\,r\big)^\ell L_k^{\ell + 1/2} \left(k_0    \Omega r^2\right)\\
        & \times \exp\left(-\frac{1}{2} k_0 \Omega r^2\right) Y_\ell^m\left(\theta, \phi\right),
    \end{split}
\end{equation}
where the normalization factor $N_{k \ell} \equiv 2^{3/4} \sqrt{2 k !/\Gamma(k + \ell + 3/2)}$ ensures that $\int |\phi_{k \ell m}|^2 \, d^3\mathbf{r} = 1$, $L_k^{\ell + 1/2}$ is the generalized Laguerre polynomial, and $Y_\ell^m$ is a spherical harmonic. The angles $\theta$ and $\phi$ are the standard spherical angles for Cartesian coordinates $\mathbf{r} \equiv (x, y, \xi)$, with the polar axis chosen to coincide with the arbitrary quantization axis $q$ (see Appendix B).

The corresponding position-space mode profile for the dispersive paraxial Hamiltonian,  $\psi_{k \ell m}(\mathbf{r}, z) \equiv \bra{\mathbf{r}}\hat{S}_\mathrm{ts} \hat{U}_\mathrm{el}(z) \ket{k \ell m}$, can be written in terms of $\phi_{k \ell m}(\mathbf{r})$ as 
\begin{equation}
    \label{eq:psiklm_wavefunc}
    \psi_{k \ell m}(\mathbf{r}, z) = \frac{1}{|\beta_2|^{1/4}}\sqrt{\frac{w_0^3}{w^3}}\exp\left(\frac{i}{2}k_0  \frac{\dot{w}}{w}\varrho^2\right)\phi_{k \ell m}\left(\frac{w_0}{w}\bm{\varrho}\right),
\end{equation}
where $\bm{\varrho} \equiv \hat{S}_\mathrm{ts} \mathbf{r} \hat{S}_\mathrm{ts}^\dag = (x, y, |\beta_2|^{-1/2} \xi)$ is the
temporally squeezed position vector. The $|\beta_2|^{-1/4}(w_0/w)^{3/2}$ coefficient is an additional normalization factor resulting from the coordinate rescalings produced by the unitary squeezing operators $\hat{S}_\mathrm{ts}$ [Eq.~\eqref{eq:S_ts}] and $\hat{S}_\mathrm{el}(z)$ [Eq.~\eqref{subeq:S_el}] and ensures the normalization $\int |\psi_{k \ell m}|^2 \, d^3\mathbf{r} = 1$. The exponential factor arises from $\hat{G}_\mathrm{el}$ [Eq.~\eqref{subeq:G_el}].

\section{Spherical Coordinates of the Harmonic Oscillator}

The position-space representation of the isotropic harmonic-oscillator eigenmodes, $\phi_{k \ell m}(\mathbf{r})$ [Eq.~\eqref{eq:phiklm_wavefunc}], depends on the spherical angles $(\theta, \phi)$ whose polar axis coincides with the arbitrary quantization axis $q$. The unit vector associated with the $q$ axis, $\mathbf{e}_q$, is restricted to lie in the $y$-$\xi$ plane, and can therefore be parameterized as $\mathbf{e}_q = \sin{\alpha_q}\,\mathbf{e}_y + \cos{\alpha_q}\,\mathbf{e}_\xi$, where $\alpha_q$ is the angle between the $q$ and $\xi$ axes. The spherical angles $(\theta, \phi)$ may then be expressed in terms of $\alpha$ and the Cartesian coordinates $(x, y, \xi)$ as
\begin{subequations}
    \begin{eqnarray}
        \begin{split}
            \theta & = \arccos\left(\frac{\mathbf{e}_q \cdot \mathbf{r}}{r}\right)\\
            & = \arccos\left(\frac{y \sin{\alpha_q} + \xi \cos{\alpha_q}}{\sqrt{x^2 + y^2 + \xi^2}}\right)
        \end{split}\\
        \begin{split}
            \phi & = \arctan\left(\frac{\mathbf{e}_x \cdot (\mathbf{r} \times \mathbf{e}_q)}{\mathbf{e}_x \cdot \mathbf{r}}\right)\\
            & = \arctan\left(\frac{y \cos{\alpha_q} - \xi \sin{\alpha_q}}{x}\right)
        \end{split}
    \end{eqnarray}
\end{subequations}

\begin{acknowledgments}
The authors would like to thank D. Chakraborty, M.R. Edwards, D. Purschke, and D. Ramsey for discussions and H. Palmer for creative assistance with the figures. The work of A.K. and J.P.P. is supported by the Office of Fusion Energy Sciences under Award Numbers DE-SC0021057, the Department of Energy National Nuclear Security Administration under Award Number DE-NA0004144, the University of Rochester, and the New York State Energy Research and Development Authority. The work of R.A. is supported by Fundação para a Ciência e Tecnologia (FCT, Portugal) grant number UI/BD/154677/2022 (DOI:10.54499/UI/BD/154677/2022).  This report was prepared as an account of work sponsored by an agency of the US Government. Neither the US Government nor any agency thereof, nor any of their employees, makes any warranty, express or implied, or assumes any legal liability or responsibility for the accuracy, completeness, or usefulness of any information, apparatus, product, or process disclosed, or represents that its use would not infringe privately owned rights. Reference herein to any specific commercial product, process, or service by trade name, trademark, manufacturer, or otherwise does not necessarily constitute or imply its endorsement, recommendation, or favoring by the US Government or any agency thereof. The views and opinions of authors expressed herein do not necessarily state or reflect those of the US Government or any agency thereof.
\end{acknowledgments}

\bibliography{bib}

\end{document}